\documentclass[aps,prb,twocolumn,a4paper,floatfix,superscriptaddress]{revtex4}
\usepackage{graphicx}
\usepackage{hyperref}
\usepackage{amsmath}
\usepackage{lineno}
\usepackage{verbatim}
\usepackage{amssymb}
\usepackage[normalem]{ulem}
\usepackage{titlesec}

\newcommand\Ueff{U_{\rm eff}}
\newcommand\UU{U_{m_{1} m_{2} m_{3} m_{4}}}
\newcommand\dd{\rm{d}}
\newcommand\rr{\mathbf{r}}

\begin{document}

\title{Self-consistent pressure-dependent on-site Coulomb correction for zero-temperature equations of state of $f$-electron metals}

\author{Bei-Lei Liu}
\affiliation{National Key Laboratory of Computational Physics, Institute of Applied Physics and Computational Mathematics, Beijing 100094, China}

\author{Yue-Chao Wang\footnote{Corresponding authors: yuechao\_wang@126.com}}
\affiliation{National Key Laboratory of Computational Physics, Institute of Applied Physics and Computational Mathematics, Beijing 100094, China}

\author{Xing-Yu Gao}
\affiliation{National Key Laboratory of Computational Physics, Institute of Applied Physics and Computational Mathematics, Beijing 100094, China}

\author{Hai-Feng Liu}
\affiliation{National Key Laboratory of Computational Physics, Institute of Applied Physics and Computational Mathematics, Beijing 100094, China}

\author{Hai-Feng Song\footnote{Corresponding authors: song\_haifeng@iapcm.ac.cn}}
\affiliation{National Key Laboratory of Computational Physics, Institute of Applied Physics and Computational Mathematics, Beijing 100094, China}

\begin{abstract}

The $f$-electron materials have many unique properties under pressure, thus of great interest in high-pressure physics and related industrial fields. However, the $f$-electrons pose a  substantial challenge to simulations since the electron correlation effects. 
In this work, we present a first-principles calculation scheme for the equations of state (EoS) of $f$-electron materials. The self-consistent pressure-dependent on-site Coulomb correction is performed based on our recently developed doubly screened Coulomb correction approach. We investigate the zero-temperature EoS over a wide range of pressures and the phase stabilities of four prototypical lanthanide and actinide metals, Pr, Eu, Th and U. 
The simulated compressive properties are in better agreement with the experimental data than those obtained by conventional density functional theory (DFT) and fixed-parameter DFT+$U$ approaches. The pressure-induced phase transitions can also be well described.  

\end{abstract}

\maketitle

%%%%%%%%%%%%%%%%%%%%%%%%%%%%%%%%%%%%%%%%%%%%%%%%%

\section{Introduction}
\label{sec:Intro}

The $f$-electron materials including the lanthanide and actinide systems, exhibit many exotic physical properties such as heavy fermions \cite{Stewart1984}, antiferromagnetic order \cite{Paschen2021} and non-Fermi-liquid behaviour \cite{Stewart2001}, of great interest in condensed matter physics and related applications \cite{Grasso2013,Benedict1994}. In particular, 
%the application of external pressure 
the compression has been observed to induce a range of promising features in these materials, including quantum phase transitions \cite{Shen2020,Maksimovic2022} and superconductivity \cite{Kruglov2018,Drozdov2019,Zhong2022}, which also provides an important mean for in-depth understanding the physical nature of $f$-electrons. 

The equations of state (EoS) 
%over a wide range of pressures 
describing the structural and compressive behaviour, plays a essential role in the study of the properties of materials with $f$-electrons under extreme conditions  \cite{Bushman1992,Dewaele2013,Liu2016,Kruglov2018,Bannon2022}. First-principles simulation is one of the most important approaches to obtain EoS data, especially at high pressures \cite{Sun2016, Rehn2021}. 
However, the $f$-electrons present a significant challenge to simulations based on density functional theory (DFT) \cite{Hohenberg1964,Kohn1965} with local density approximation or generalized gradient approximation, mainly attributed to the electron correlation effects.
 
Much effort has been devoted to the development of suitable methods for $f$-electron materials.
For lanthanides, a simplified approach is treating the $4f$-electrons as core-electrons \cite{Eriksson1995,Delin1998-1,Delin1998-2}. This approach provides a reasonable binding properties at ambient pressure. However, its inherent disadvantage lies that it cannot properly describe the delocalization of $f$-electron under pressure \cite{Soderlind2014-La}. In addition, the treatment of $f$-electrons as core-level is not applicable to light actinides, whose $5f$-electron participates more actively in chemical bonding \cite{Johansson1995,Adak2011}. In recent years, the on-site Coulomb correction of DFT in static mean-field approximation DFT+$U$ \cite{Mohanta2010, Amadon2018}, DFT plus dynamical mean-field theory (DMFT) \cite{Locht2016, Amadon2016}, self-interaction corrected (SIC) forms \cite{Strange1999,Petit2010} and DFT with orbital polarization (DFT+OP) \cite{Soderlind2004,Soderlind2014-La}, have been developed for $f$-electron materials. 
Among them, the method in combination with model Hamiltonian correction (such as DFT+$U$ and DFT+DMFT) has become one of the most widely used and powerful approaches to tackle correlated $f$-systems. However, its on-site Coulomb interaction parameters, which have a significant influence on the performance, are typically unknown in advance. These parameters are strongly depend on the behaviour of the electron, and thus generally different for different compositions and structures. 
In particular, it has been observed that better performance in EoS calculations can be achieved by tuning the on-site Coulomb interaction strengths at different pressure intervals \cite{Verma2013,Mondal2017,Sun2020}.
However, determining the on-site Coulomb interaction parameters is a challenging task \cite{Cococcioni2005,Aryasetiawan2004}. Recently, the doubly screened Coulomb correction (DSCC) approach \cite{Liu2023} makes a favorable balance between accuracy and computational efficiency, providing a practical way to simulate the on-site Coulomb interaction parameters.

In this work, we propose a computational scheme for the EoS over a wide range of pressures for $f$-electron materials, named DFT+DSCC. In this scheme, the $f$-electron is corrected by the static mean-field Hubbard model, which is similar to the traditional DFT+$U$ method. However, the on-site Coulomb interaction parameters are adjusted according to the $f$-electron state adaptively, based on our recently developed DSCC approach. In order to demonstrate the applicability of the proposed scheme, tests were conducted on four lanthanide and actinide materials: Pr, Eu, Th and U. We compare the results of this scheme with those of DFT and fixed-parameter DFT+$U$ approaches. The present scheme yields a more accurate description of zero-temperature EoS over a wide range of pressures. In addition to EoS, this scheme is able to give a reasonable description of the phase stabilities.

The paper is organized as follows: In Sec. II, we introduce the computational scheme and related details. In Sec. III, the calculated results of EoS over a wide range of pressures and phase stabilities are presented, and compared with accessible experimental and related theoretical data. Finally, we conclude in Sec. IV.

%%%%%%%%%%%%%%%%%%%%%%%%%%%%%%%%%%%%%%%%%%%%%%%%%

\section{Method and Computational Details}
\label{sec:Meth}

In the DFT+DSCC scheme, a static mean-field %approximation to the 
Hubbard model \cite{Anisimov2010} is used to correct $f$-electrons. The ground-state energy can be written as follows: 
\begin{equation}
\label{eq:tot}
E_{\rm tot} = E_{\rm DFT} + E_{\rm Hub} + E_{\rm dc},
\end{equation}
where $E_{\rm DFT}$ refer to the energy of system described by DFT, the double counting term $E_{\rm dc}$ is used to cancel the electron-electron interaction that have already been included in DFT. $E_{\rm Hub}$ described the interaction energy between $f$-electrons form Hubbard model, it takes the following form:
\begin{equation}
\begin{split}
\label{eq:on-site}
&E_{\rm Hub} =\frac{1}{2}\sum_{{m},\sigma}\big[\UU n_{m_1,m_2}^\sigma n_{m_3,m_4}^{\bar{\sigma}}  \\
& +(\UU-U_{m_{1} m_{3}m_{4}m_{2}})n_{m_1,m_2}^\sigma n_{m_3,m_4}^\sigma\big],
\end{split}
\end{equation}
where $\left\{m\right\}$ is the index set of local orbitals, $m_1,m_2,m_3,m_4\in\{m\}$, $\sigma$ stands for the spin, $n_{i,j}^\sigma$ is the occupation matrix element, which are determined by the projection of occupied Kohn-Sham orbitals onto local orbitals \cite{Shick1999,Bengone2000,Amadon2008}. 

The DSCC approach is adopting to construct the Hubbard model in our scheme. The interaction matrix element is calculated by
\begin{equation}
\label{eq:int}
U_{ijkl}=\int\int{\varphi_i^\ast(\rr)\varphi_j^\ast(\rr')v_{\rm sc}(\rr,\rr')\varphi_k{(\rr)\varphi}_l(\rr')\dd\rr \dd\rr'},
\end{equation}
where the screened Coulomb potential $v_{\rm sc}$ is determined by the Fourier transformation into real space of the doubly screened model dielectric function \cite{Cappellini1993,Liu2023}. 
The local orbitals $\varphi_{m}$ is constructed as the atomic orbitals of $f$-shell ($l$=3) \cite{Wang2019, Novak2006}.
Then the ground-state properties can be calculated in this framework.
In order to obtain the EoS, we directly fit the pressure ($P$) and volume ($V$) because the energies are generally not directly comparable at different $U,J$ values. In contrast to the conventional DFT+$U$ approach calculating the EoS over a wide range of pressures based on a fixed, semi-empirical $U,J$ values (we denoted as DFT+$U_{\rm fix}$ hereafter), our scheme evaluates the on-site Coulomb interaction strength in a pressure-dependent and self-consistent way. Thus this scheme allows for a more accurate description of the phenomenon induced by localization-delocalization transition of $f$-electron under pressure, as well as the treatment of materials with quite weak correlation effect.

Two lanthanide systems praseodymium (Pr), europium (Eu) and two actinide systems thorium (Th), uranium (U) are selected to test the applicability of our calculation scheme. The $4f$-electrons are generally considered to be delocalized under pressure \cite{Benedict1986,Bannon2022}. In the lanthanide series, Pr shows a pronounced phase transition related to the $f$-electron delocalization under pressure \cite{Soderlind2002-2,Bannon2022}, whereas this tendency is weaker for Eu, whose $f$-electron remains localized at quite high pressure with strong local magnetic moment \cite{Min1986-2,Bi2016}. The actinides show an increase in localization along the series \cite{Moore2009}. The $5f$-electrons of the light actinides Th and U are itinerant and contribute to the bonding \cite{Skriver1980,Johansson1995,Bouchet2017}, unlike the heavy actinides, where the behaviour of the $f$-electron is similar to the $f$-state in the lanthanides \cite{Benedict1986,Verma2013}. Additionally, there is an abundance of high-pressure experimental and theoretical data available for these systems to compare with. 

At low pressure, Pr exhibit some high symmetry structures as the double hexagonal close-packed (dhcp), face-centered cubic (fcc), distorted-fcc and body-centered orthorhombic (bco) in succession \cite{Baer2003,Bannon2022}. Above $\sim$20 GPa, Pr transforms to a low-symmetry $\alpha$-U structure, with a volume collapse. The $\alpha$-U structure is then stabilized up to 185 GPa. In this work, we select the representative fcc and $\alpha$-U structures in the low and high pressure ranges, respectively, for simulation.   
Eu undergoes a bcc-hcp phase transition at 12 GPa \cite{Takemura1985,Bi2011}. The hcp phase remains stable until 30 GPa, after which incommensurately modulated monoclinic crystal structure with symmetry of $C2/c$ symmetry is observed, until it transforms into an orthorhombic ($Pnma$) structure at 78 GPa. We primarily focus on the bcc and hcp structures of Eu, and the phase transition between them.
The investigation of the incommensurate Eu structure is beyond the scope of the present work.
The involvement of the $5f$ states in the chemical bonds of Th leads to a ground-state fcc structure \cite{Johansson1995}. In the range of 70-100 GPa, Vohra and Akella reported a fcc-bct phase transition \cite{Armstrong1959,Vohra1991}. The bct phase is observed to be stable under high pressure, up to 300 GPa. In terms of U, it crystallizes into the $\alpha$-U phase under ambient conditions. Experimental evidence has demonstrated that $\alpha$-U remains the most stable phase at least 100 GPa \cite{Akella1990,Yoo1998,LeBihan2003,Dewaele2013}. In recent years, several theoretical studies have postulated the transition of $\alpha$-U to the bct phase at $\sim$270 GPa \cite{Adak2011,Kruglov2019,Pan2024}. However, there are no documented experimental observations of U at such high pressure.

All the calculations are performed in the Vienna ${ab~initio}$ simulation package (VASP) \cite{Kresse1996,Kresse1999}, which is based on projector-augmented wave (PAW) farmework \cite{Blochl1994,Kresse1999,Fang2019}. The exchange and correlation functional is given by the GGA of Perdew-Burke-Ernzerhof \cite{PBE1996} (PBE) form.  
The cut-off energy of 500 eV for the plane-wave basis. The ${\bf k}$-point grids of Brillouin zone sampling based on the Monkhorst-Pack scheme \cite{Monkhorst1976} of $8\times 8 \times 8$ for $\alpha$-U structure, and $14\times 14 \times 14$ for other high symmetry structures were applied. In the calculation of EoS, structural relaxations were performed by relaxing lattice parameters and internal coordinates at fixed volume to determine the hydrostatic ground states. In order to calculate the enthalpy difference at the same pressure, structural relaxations at prescribed pressure also were performed for different phases. In structure relaxation, the improved gradient descent algorithms \cite{Hu2022,Hu2024} are used at fixed volume and pressure, respectively. The convergence criterion of forces acting on atoms were set as 0.01 eV/\r{A}. In static calculations, the 1st order Methfessel-Paxton method \cite{Methfessel1989} was used with the parameter $\sigma$ = 0.05 eV, and the convergence criterion of total energy were set as $1\times10^{-6}$ eV. 
All the DFT+$U_{\rm fix}$ calculations were performed using the simplified formulation proposed by Dudarev $et~al$ \cite{Dudarev1998},  which depends on the effective on-site Coulomb correction parameter $\Ueff \equiv U-J$ only. For DFT+$U_{\rm fix}$ calculations, we set $\Ueff$ = 5.0 eV for Pr and Eu, and 2.0 eV for Th and U, which is in the commonly used range for light lanthanides and light actinides \cite{Soderlind2014-La,Locht2016,Qiu2020}. The fully localized limit form \cite{Anisimov1993} was employed in order to account for the double counting term in the Hubbard correction. The spin-orbit coupling (SOC) is included for for all test systems. 

%%%%%%%%%%%%%%%%%%%%%%%%%%%%%%%%%%%%%%%%%%%%%%%%%

\section{Results and Discussion}
\label{sec:Res}

\subsection{Praseodymium}

\begin{figure}[tbp]
\centering
\includegraphics[width=0.49\textwidth]{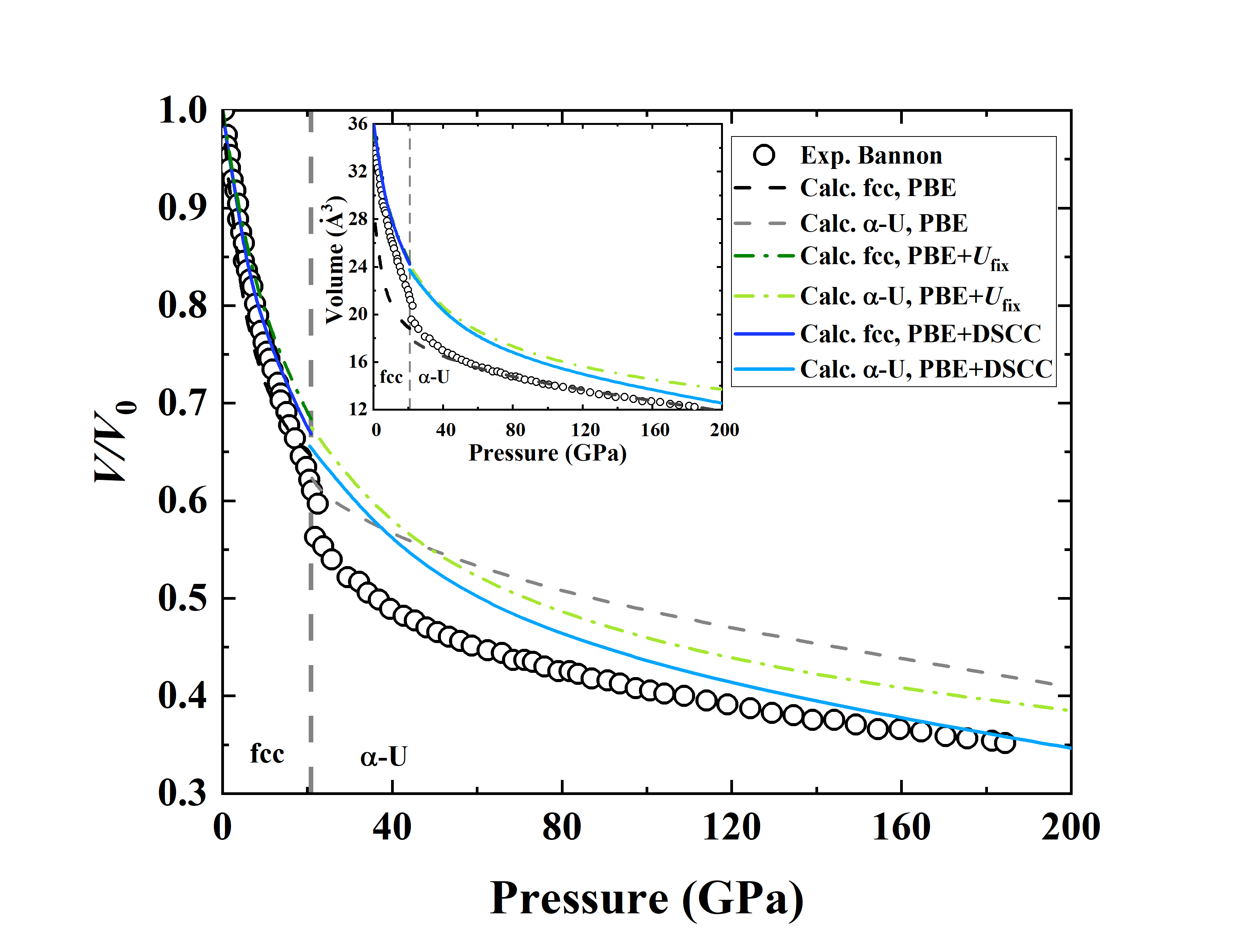}
\caption{The static EoS $P$-$V/V_0$ of Pr calculated by PBE, PBE+$U_{\rm fix}$, and PBE+DSCC. Inset shows the corresponding $P$-$V$ relationships. Experimental data also present for comparison. }\label{fig:Pr-eos}
\end{figure}

\begin{table}[tbp]
\caption{Structural parameters $V_0$, $B_0$ and $B'_0$ of Pr by experimental and theoretical approaches.{\label{tab:struc:Pr} }}
\begin{tabular*}{0.49\textwidth}{@{\extracolsep{\fill}} l|cccc}
\hline\hline
     &$V_0$ (\r{A}$^3$)          &$B_0$ (GPa)      &$B'_0$ \\
 \hline
Exp. Baer \cite{Baer2003}        &34.2     &25.1    &2.9        \\
Exp. O'Bannon \cite{Bannon2022}       &34.7     &25.0     &2.8  \\
PBE    &25.947     &18.849     &7.774          \\
PBE+$U_{\rm fix}$($\Ueff$=5 eV)   &35.598     &30.671     &2.919          \\
PBE+DSCC  &36.240     &23.989     &3.846          \\
\hline\hline
\end{tabular*}
\end{table}

We first test the performance of self-consistent on-site Coulomb correction scheme on the lanthanide metal Pr. 
Fig.\ref{fig:Pr-eos} shows the calculated curve of the reduced volume $V/V_0$ versus the pressure $P$. The inset shows the curve of volume $V$ versus the pressure $P$. We also provide a comparison with PBE, conventional PBE+$U_{\rm fix}$ using fixed on-site Coulomb interaction parameters, and experimental data. 

PBE remarkably underestimates the equilibrium volume at 0 GPa, with a discrepancy of approximately 20\% (see also Table \ref{tab:struc:Pr}, the fitted third-order Birch-Murnaghan EoS parameters). and underestimates the volume throughout the low-pressure range. However, above $\sim$40 GPa, PBE produces the closest results in comparison to the experimental data.
PBE+$U_{\rm fix}$ provides a notable improvement in the structural property at low pressures. The discrepancy between the calculated and experimental volume is at most 9\% within 0–15 GPa. Nevertheless, at higher pressures, an increased divergence is observed. Up to $\sim$185 GPa, the discrepancy between the calculated and experimental volume is 14.5\%.
The $P$-$V$ curve calculated by PBE+DSCC shows an obvious transition in the range of 20 to 200 GPa, moving from a closer alignment with that of PBE+$U_{\rm fix}$ to a more similar behaviour to that of PBE.
Due to the satisfactory description of the structural behaviour both at atmospheric pressure and at pressures of hundreds of GPa, PBE+DSCC has a good performance on the $P$-$V/V_0$ curve, with a deviation of less than 5\% over the range 130-185 GPa, where $V_0$ is set as the equilibrium volume at 0 GPa calculated by each method.
It should be noted that the $P$-$V/V_0$ curve reflects the compressive nature of materials, and is widely used as a standard for equation of state modelling \cite{Wang2000-2,Li2001,Adak2011,Tian2024}. It should be noted that the calculated data presented in this work are all at 0 K. For the $P$-$V/V_0$ curves, the 300 K temperature effect is generally less than 1\%. 

\begin{figure}[tbp]
\centering
\includegraphics[width=0.49\textwidth]{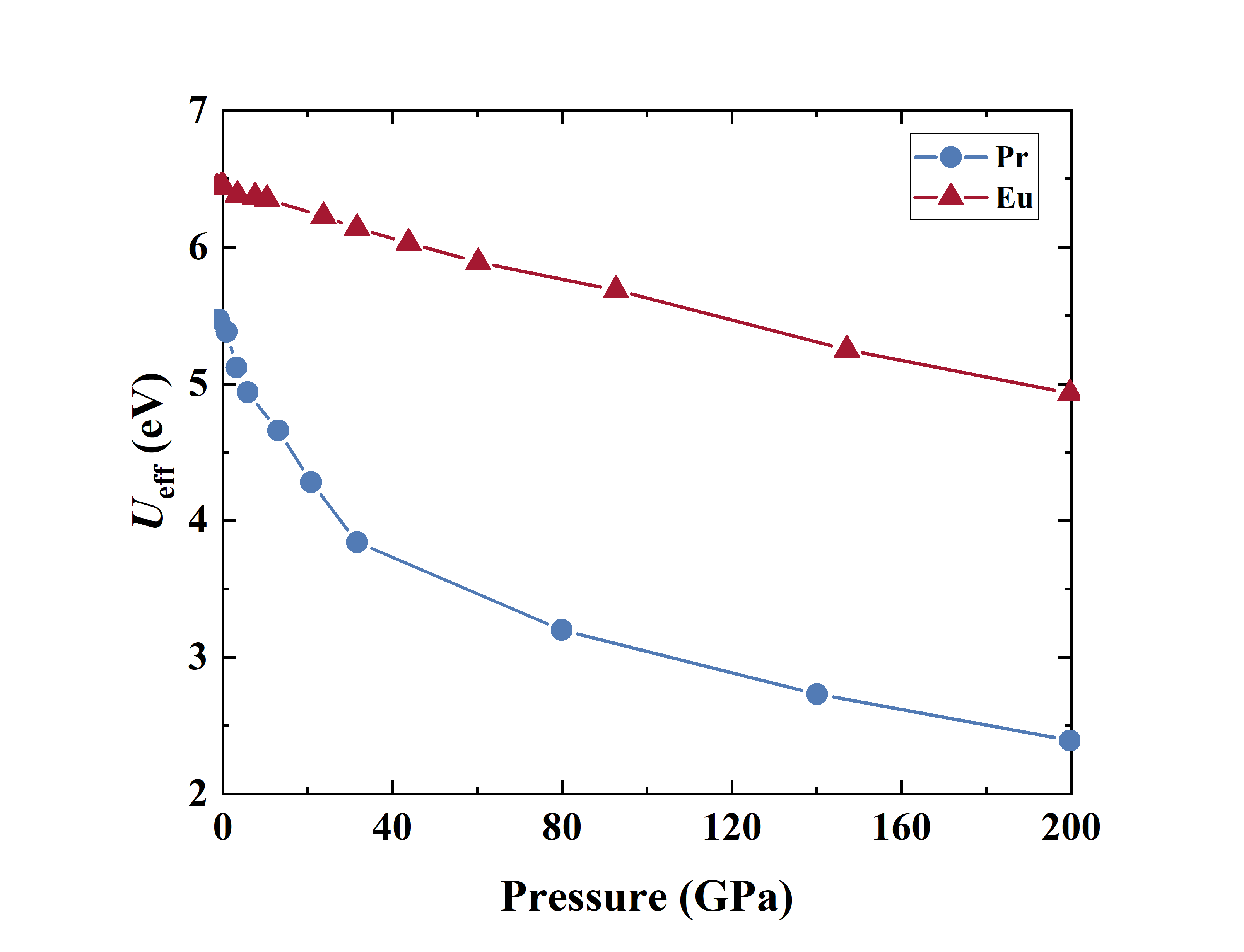}
\caption{The effective Coulomb interaction strength $U_{\rm eff}$ of Pr and Eu as a function of pressure $P$ calculated by DSCC approach.}\label{fig:Pr-ueff}
\end{figure}

The feature of our calculation scheme is that it takes into account the pressure-dependent on-site Coulomb interaction parameters $U,J$. Fig.\ref{fig:Pr-ueff} shows the calculated effective on-site Coulomb interaction strength $U_{\rm eff} \equiv U-J$ versus pressure of Pr.  
The $U_{\rm eff}$ calculated by DSCC is 5.41 eV at 0 GPa, which is in close agreement with the value of 5.5 eV estimated by Lang $et~al.$ \cite{Lang1981} using X-ray photoelectron spectroscopy (XPS) and bremsstrahlung isochromat spectroscopy (BIS) data. It is also consistent with the previous theoretical results within the range of 4.0-6.0 eV \cite{Min1986,Nilsson2013,Liu2023-1}. The reduction in $\Ueff$ is observed with increasing pressures, albeit at a progressively slower rate. It decreases by approximately 44\% at 100 GPa, and 55\% at 200 GPa.

\begin{figure}[tbp]
\centering
\includegraphics[width=0.49\textwidth]{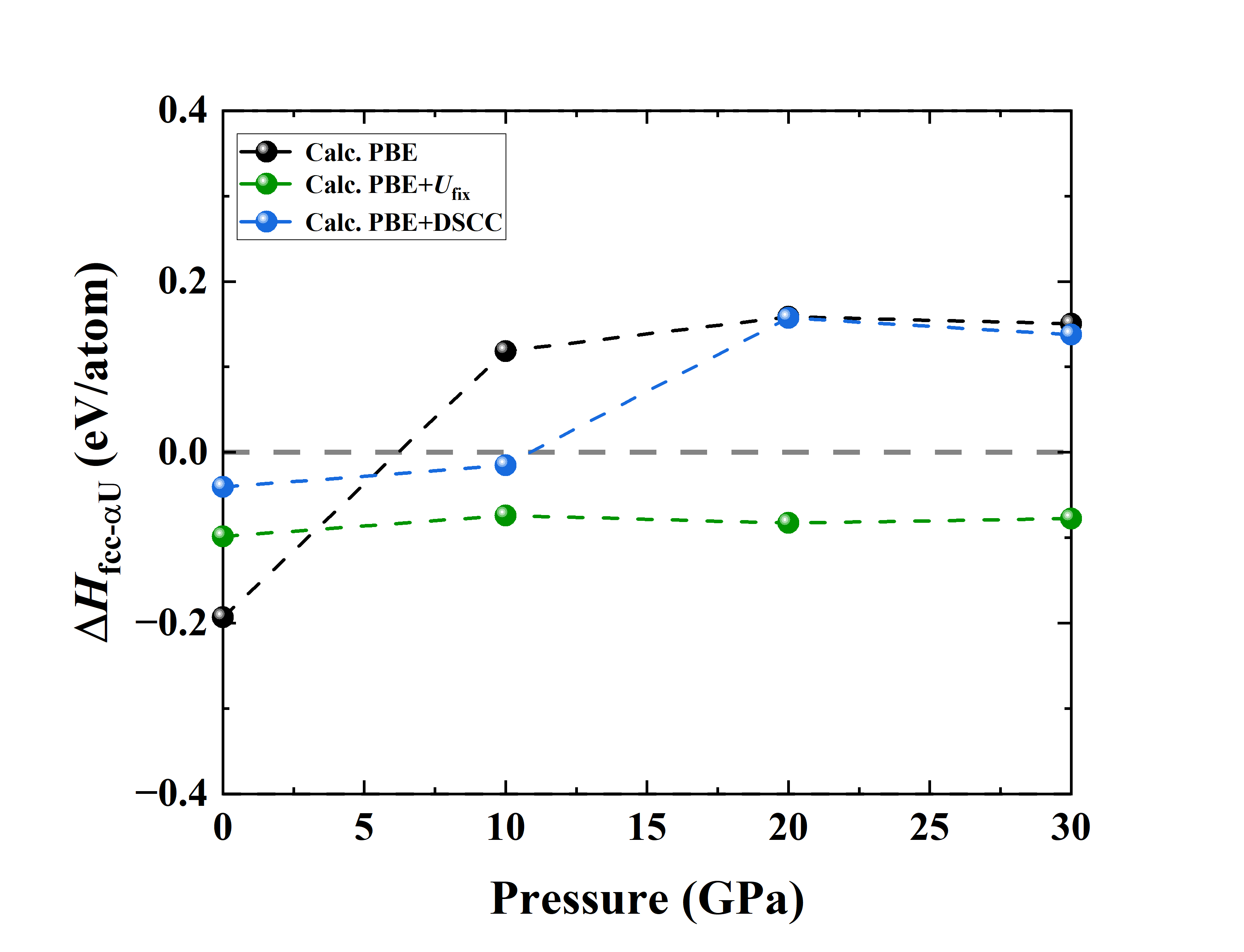}
\caption{The enthalpy difference of fcc structure related to $\alpha$-U phase of Pr as a function of pressure, calculated by PBE, PBE+$U_{\rm fix}$, and PBE+DSCC.}\label{fig:Pr-dh}
\end{figure}

Experiments indicate that there is a transition from high to low symmetry structure in Pr, which is believed closely linked to the delocalization of 4$f$ electron \cite{Soderlind2002-2,Bannon2022}. 
At about 20 GPa, Pr undergoes this phase transition with a $\sim$9\% volume collapse. The volume collapse described by PBE, PBE+$U_{\rm fix}$ and PBE+DSCC are $\sim$4\%, $\sim$2\% and $\sim$2.5\% at this pressure, respectively.
In order to investigate this phase transition, we also calculate the enthalpy differences of fcc structure and $\alpha$-U phase by PBE, PBE+$U_{\rm fix}$, and PBE+DSCC, as shown in Fig.\ref{fig:Pr-dh}. PBE predicts a reversal of the enthalpy of fcc and $\alpha$-U at $\sim$7 GPa. The enthalpy of the high-symmetry structure fcc calculated by PBE+$U_{\rm fix}$ is lower than $\alpha$-U continually. PBE+DSCC predicts a reversal at $\sim$11 GPa. It should be noted that Pr exhibit as d-fcc, bco structure at 12-20 GPa, so the reversal between fcc and $\alpha$-U enthalpies cannot directly comparable to the experimental transition pressure. Nevertheless, a continually lower fcc enthalpy predicted by PBE+$U_{\rm fix}$ is not reasonable.

\subsection{Europium}

\begin{figure}[tbp]
\centering
\includegraphics[width=0.49\textwidth]{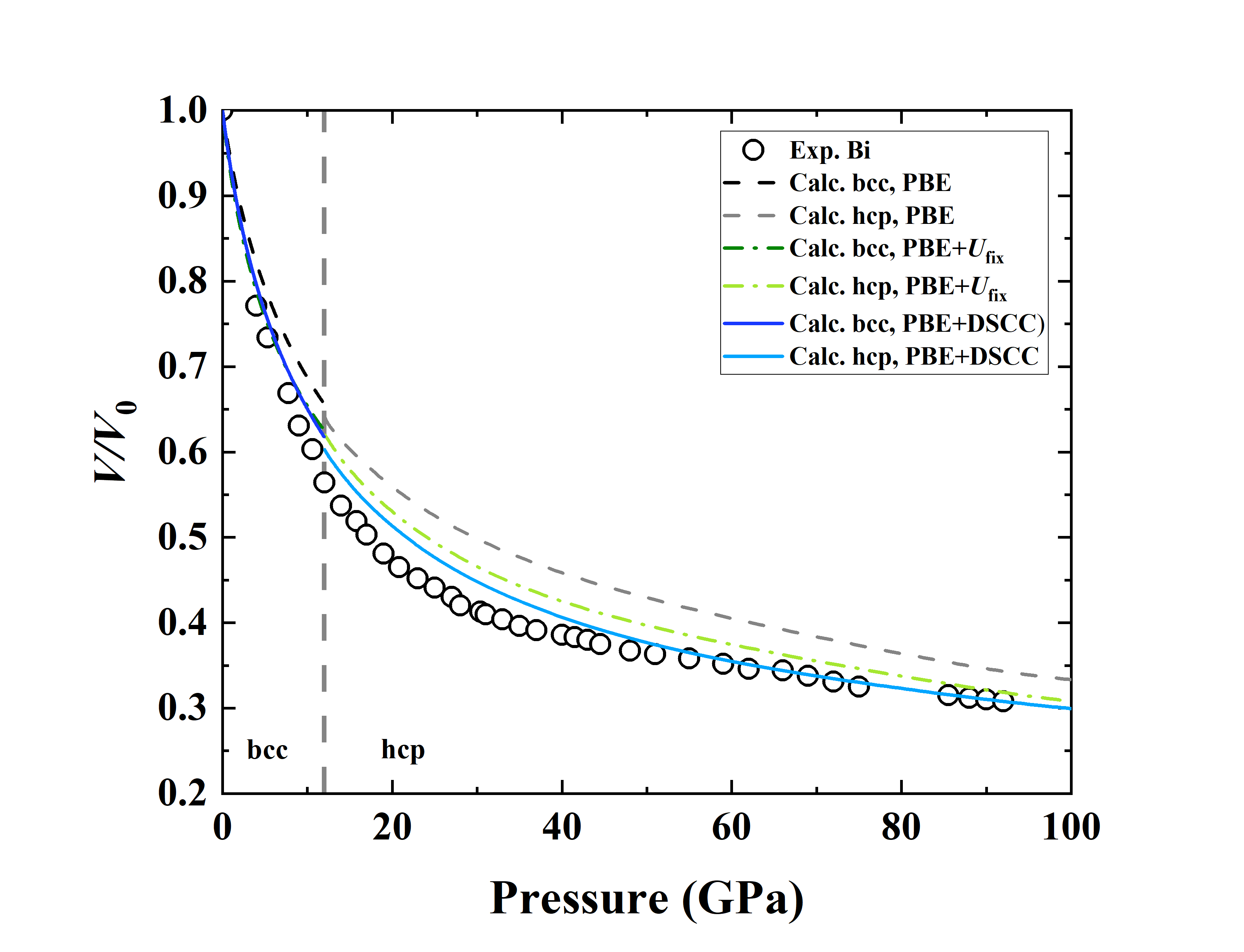}
\caption{The static EoS $P$-$V/V_0$ of Eu calculated by PBE, PBE+$U_{\rm fix}$, and PBE+DSCC. Experimental data also present for comparison.}\label{fig:Eu-eos}
\end{figure}

\begin{table}[tbp]
\caption{Structural parameters $V_0$, $B_0$ and $B'_0$ of Eu by experimental and theoretical approaches.{\label{tab:struc:Eu} }}
\begin{tabular*}{0.45\textwidth}{@{\extracolsep{\fill}} l|cccc}
\hline\hline
     &$V_0$ (\r{A}$^3$)          &$B_0$ (GPa)      &$B'_0$ \\
 \hline
Exp. Takemura \cite{Takemura1985}       &48.0     &11.7     &3.0 \\
Exp. Bi \cite{Bi2011}        &47.95     &10.9    &3.0        \\
PBE    &43.208     &14.835     &3.276          \\
PBE+$U_{\rm fix}$($\Ueff$=5 eV)   &47.641     &10.910     &3.679          \\
PBE+DSCC  &48.855     &12.341     &3.177          \\
\hline\hline
\end{tabular*}
\end{table}

Next, we show the simulated results for Eu, which is also a lanthanide metal. 
As shown in Table \ref{tab:struc:Eu}, PBE underestimates the equilibrium volume of Eu at 0 GPa by about 10\%. By introducing the on-site Coulomb correction, PBE+$U_{\rm fix}$ and PBE+DSCC reproduces the equilibrium volume with a deviation less than 2\%.
Fig.\ref{fig:Eu-eos} shows that the calculated compression curves and experimental data. The result calculated by PBE diverge considerably from the experimental data. Even with the incorporation of $U$, PBE+$U_{\rm fix}$ still exhibit a divergence of 10\% at 40 GPa. In contrast, PBE+DSCC scheme demonstrates an error of less than 5\% in comparison to the experimental values at 40 GPa, and the discrepancy between the calculated and experimental results diminishes under pressure. 

The calculated $U_{\rm eff}$ by DSCC of Eu is presented in Fig.\ref{fig:Pr-ueff}, and compared with those of Pr. At 0 GPa, $U_{\rm eff} = 6.44$ eV. This result is consistent with previous theoretical pragmatic $U_{\rm eff}$ values in the range of 5-7 eV \cite{Locht2016,Liu2023-1}. It should be noted that previous results also gave $U_{\rm eff}$ values of about 9-10 eV or even higher \cite{Lang1981,Min1986}. However, these values are obviously out of the practical range in the simulations. In terms of pressure evolution, the $U_{\rm eff}$ values for Eu decrease almost linearly. $U_{\rm eff}$ decreases by approximately 12.5\% at 100 GPa. The decrease in the on-site Coulomb interaction of Eu under compression is apparently less pronounced than that observed for Pr. This is consistent with the extent of $f$-electron delocalization transition for Eu and Pr. For Pr, the fcc to $\alpha$-U phase transition at $\sim$20 GPa is thought to be closely related to the $f$-electron delocalization\cite{Bannon2022,Soderlind2002-2}. However, for Eu, a strong localized magnetic moment persists at quite high pressures \cite{Bi2016}.

\begin{figure}[tbp]
\centering
\includegraphics[width=0.49\textwidth]{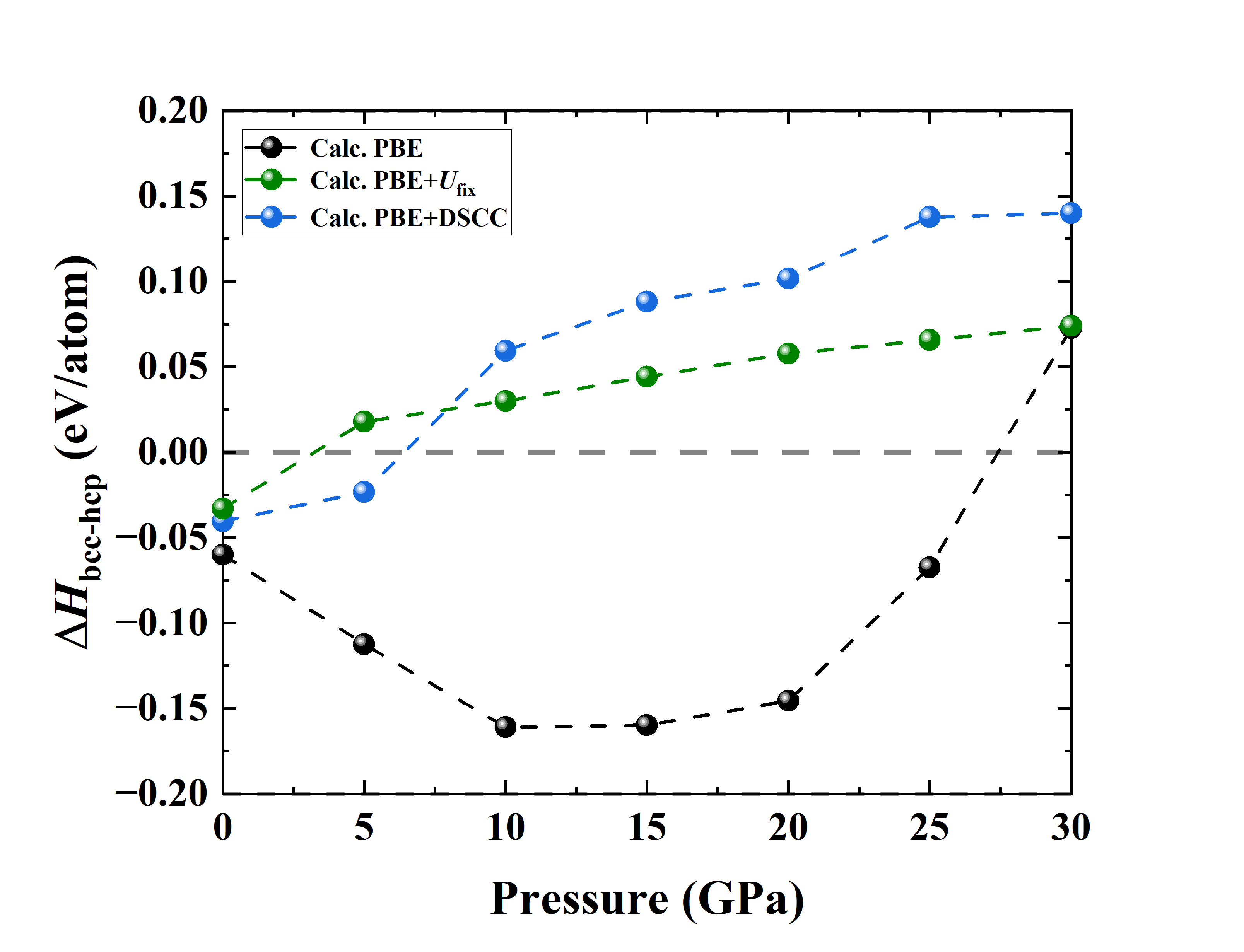}
\caption{The enthalpy difference of bcc structure related to hcp structure of Eu as a function of pressure, calculated by PBE, PBE+$U_{\rm fix}$, and PBE+DSCC.}\label{fig:Eu-dh}
\end{figure}

Transition from bcc to hcp structure in Eu at about 12 GPa, accompanied with a volume collapse at about 3\% \cite{Takemura1985,Bi2011}. All these methods able to describe this collapse, with about 2.25\% for PBE and PBE+DSCC, but 1.5\% for PBE+$U_{\rm fix}$. Fig.\ref{fig:Eu-dh} shows the calculated results of bcc-hcp enthalpy difference by PBE, PBE+$U_{\rm fix}$ and PBE+DSCC. The PBE+DSCC predicts a phase transition of $\sim$7.5 GPa. In comparison, the bcc-hcp enthalpy reversal for PBE occurs too late, at $\sim$27.5 GPa, while the reversal for PBE+$U_{\rm fix}$ occurs too early, at $\sim$3.5 GPa. 

The calculated static EoS and phase stability of lanthanide metals Pr and Eu are presented above. It is notable that the behaviour of the $4f$ electron is markedly sensitive to pressure. The inclusion of an on-site Coulomb correction can facilitate a more accurate description of structural properties in the vicinity of atmospheric pressure. However, PBE+$U_{\rm fix}$ systematically overestimates the volume at high pressure, primarily due to its disadvantage in capturing the delocalization of $4f$-electron under pressure \cite{Soderlind2014-La}. PBE+$U_{\rm fix}$ also unable to reproduce the known phase transition of Pr form high-symmetry structure to $\alpha$-U phase. The PBE+DSCC scheme achieving a good correction for the equilibrium volume at 0 GPa while better describing the high pressure behaviour compared to the PBE+$U_{\rm fix}$. Nevertheless, in the intermediate range accompanying the volume-collapse phase transition, the performance of PBE+DSCC in comparison to the experimental data remains unsatisfactory. We speculate that the origin of this phenomenon may be related to the fact that DSCC still treats $f$-electrons in the static mean-field approximation, which does not allow a good description of the change in $f$-$spd$ hybridization \cite{Amadon2015}. To address this issue, we will continue to improve the computational scheme in future work. 

\subsection{Uranium}

\begin{figure}[tbp]
\centering
\includegraphics[width=0.49\textwidth]{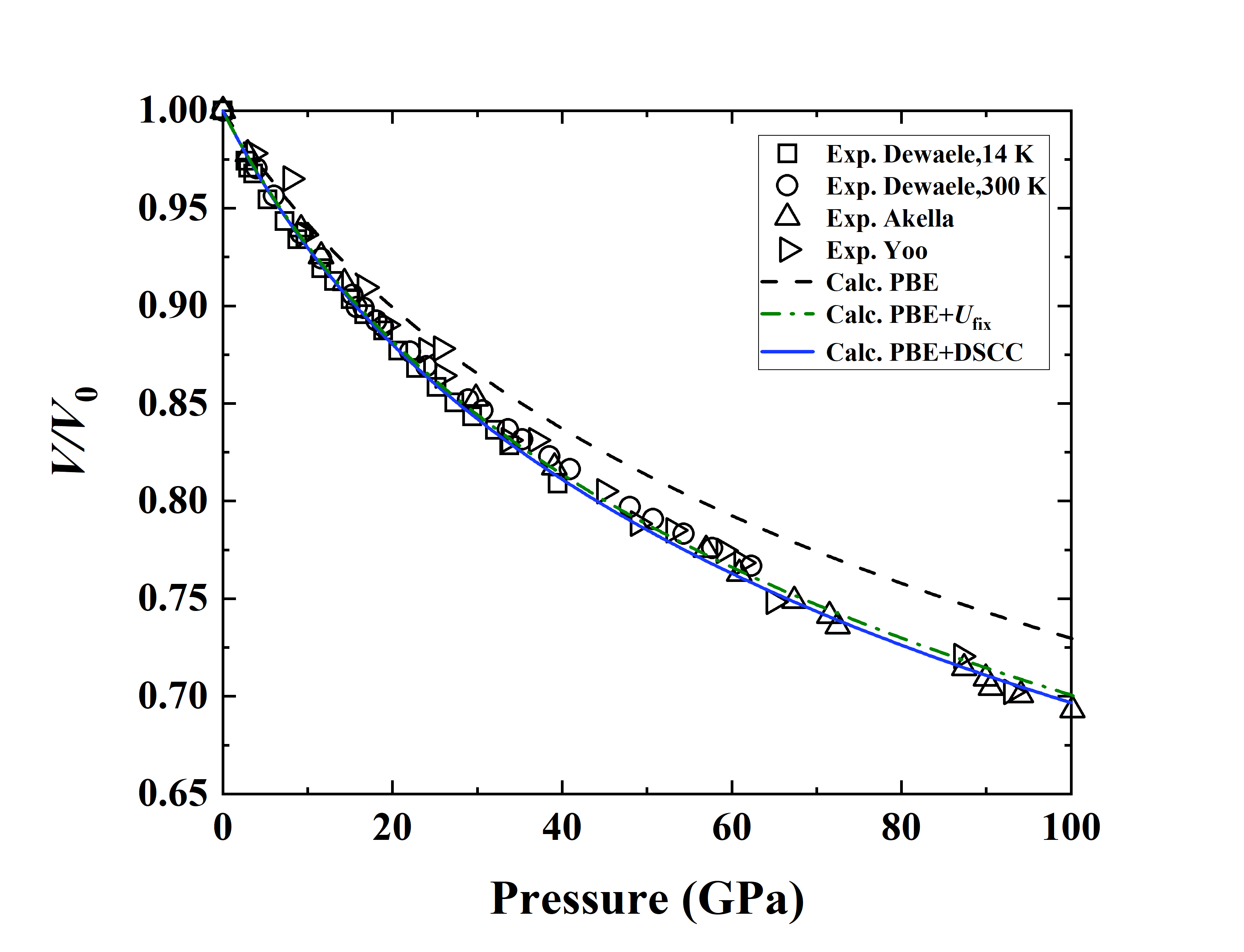}
\caption{The static EoS $P$-$V/V_0$ of U calculated by PBE, PBE+$U_{\rm fix}$, and PBE+DSCC. Experimental data also present for comparison.}\label{fig:U-eos}
\end{figure}

\begin{table}[tbp]
\caption{Structural parameters $V_0$, $B_0$ and $B'_0$ of U by various experimental and theoretical approaches.{\label{tab:struc:U} }}
\begin{tabular*}{0.45\textwidth}{@{\extracolsep{\fill}} l|cccc}
\hline\hline
     &$V_0$ (\r{A}$^3$)          &$B_0$ (GPa)      &$B'_0$ \\
 \hline
Exp. Akella  \cite{Akella1990}        &--     &138.7     &3.78  \\
Exp. Yoo  \cite{Yoo1998}     &--    &135.5     &3.8             \\
Exp. Le Bihan \cite{LeBihan2003}    &20.77     &104    &6.22   \\
Exp. Dewaele \cite{Dewaele2013}       &20.66     &114.5     &5.46  \\
PBE  &20.15     &138.80     &5.075          \\
PBE+$U_{\rm fix}$($\Ueff$=2 eV)  &21.71     &116.95     &4.881          \\
PBE+DSCC  &21.61     &115.49     &4.784          \\
\hline\hline
\end{tabular*}
\end{table}

The actinide metal uranium (U) has significant applications in nuclear science and technology. Although the available evidence suggests that electronic correlation effect play an important role in modelling the properties of intermetallic compounds of U \cite{Shick2001,Yaresko2003,Antonov2003,Antonov2003-2,Zhang2014,Xu2019,Wu2022,Broyles2023}, there is no consensus on the necessity of including the on-site Coulomb correlation for the simulation of U itself \cite{Xiong2013,Xie2013,Xie2016,Soderlind2014,Soderlind2014-2}. Table \ref{tab:struc:U} presents the structural parameters for different computational schemes, with a comparison to previous experimental data. 
PBE is about 3\% too small for the equilibrium volume at 0 GPa. PBE+$U_{\rm fix}$ and PBE+DSCC overestimates the equilibrium volume by $\sim$5\%. 
Fig.\ref{fig:U-eos} shows the compression curves. The comparison indicates that the correlation corrections have a substantial impact on the compression property. PBE+DSCC and PBE+$U_{\rm fix}$ obviously improves the accuracy of the bulk modulus.
According to Le Bihan $et~al.$\cite{LeBihan2003}, the early bulk moduli measured by X-ray are overestimated due to non-hydrostatic stress effects (such as 138.7 GPa obtained by Akella $et~al.$ \cite{Akella1990}). Recent X-ray study using quasihydrostatic media helium and properly calibrated pressure gauge have yielded values of the bulk modulus that were closer to those of other experimental techniques, with a value of about 110 GPa \cite{Dewaele2013}. PBE overestimates the bulk modulus by about 25\% and the bulk moduli obtained by PBE+DSCC and PBE+$U_{\rm fix}$ are very close to the experimental data.

\begin{figure}[tbp]
\centering
\includegraphics[width=0.49\textwidth]{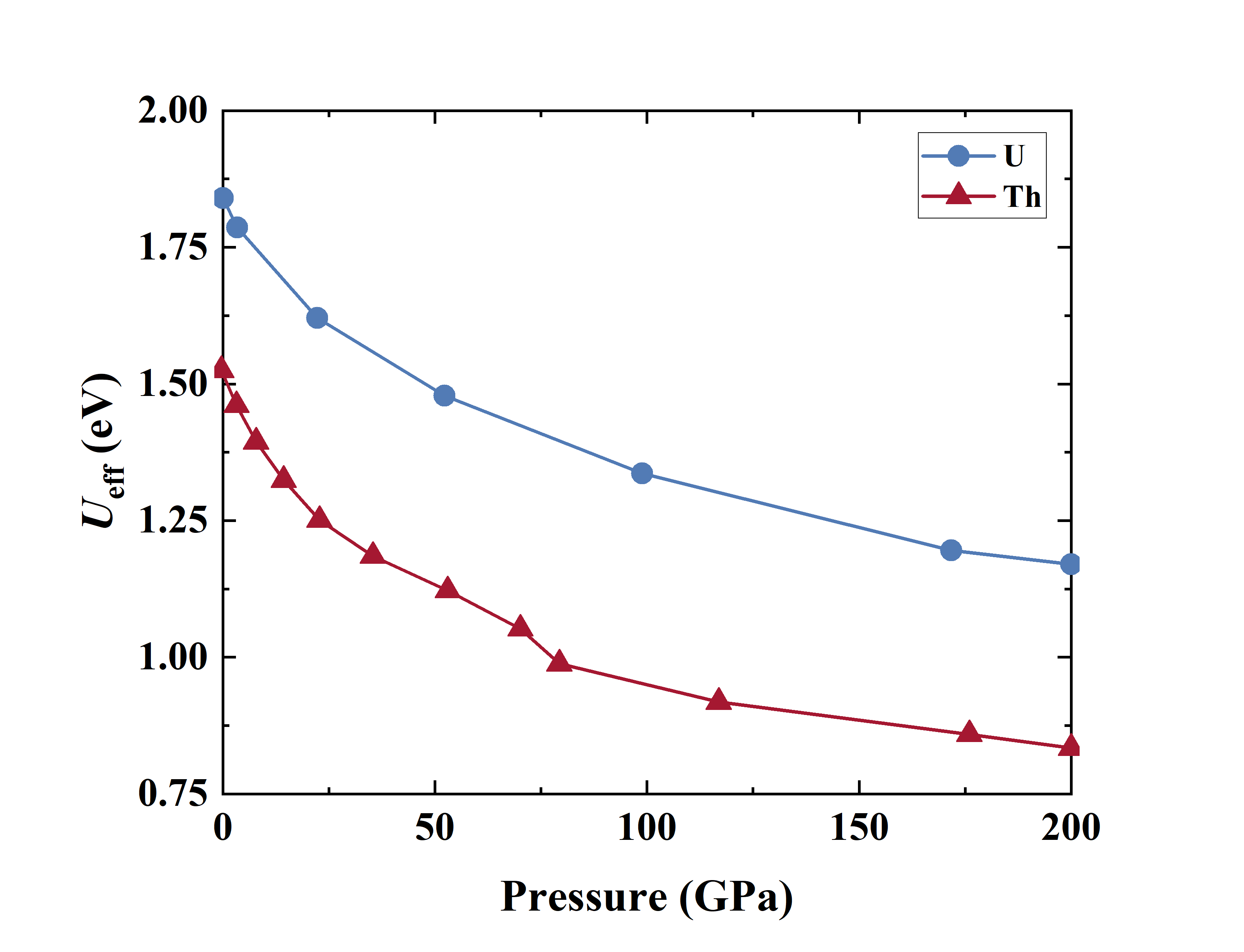}
\caption{The effective Coulomb interaction strength $U_{\rm eff}$ of U and Th as a function of pressure $P$ calculated by DSCC approach.}\label{fig:U-ueff}
\end{figure}

The calculated $U_{\rm eff}$ of U are shown in Fig \ref{fig:U-ueff}. The $\Ueff$ obtained by DSCC at 0 GPa is 1.84 eV, which is close to the theoretical values of 1.87 eV and 2.4 eV obtained in the previous calculation using linear response approach \cite{Xie2013,Qiu2020}, but is significantly larger than cRPA value $\Ueff$ = 0.4 eV \cite{Amadon2016}. In terms of the trend with pressure, the the effective on-site Coulomb interaction strength gradually decreases with increasing pressure, and is reduced by about 40\% at 200 GPa.

\begin{figure}[tbp]
\centering
\includegraphics[width=0.49\textwidth]{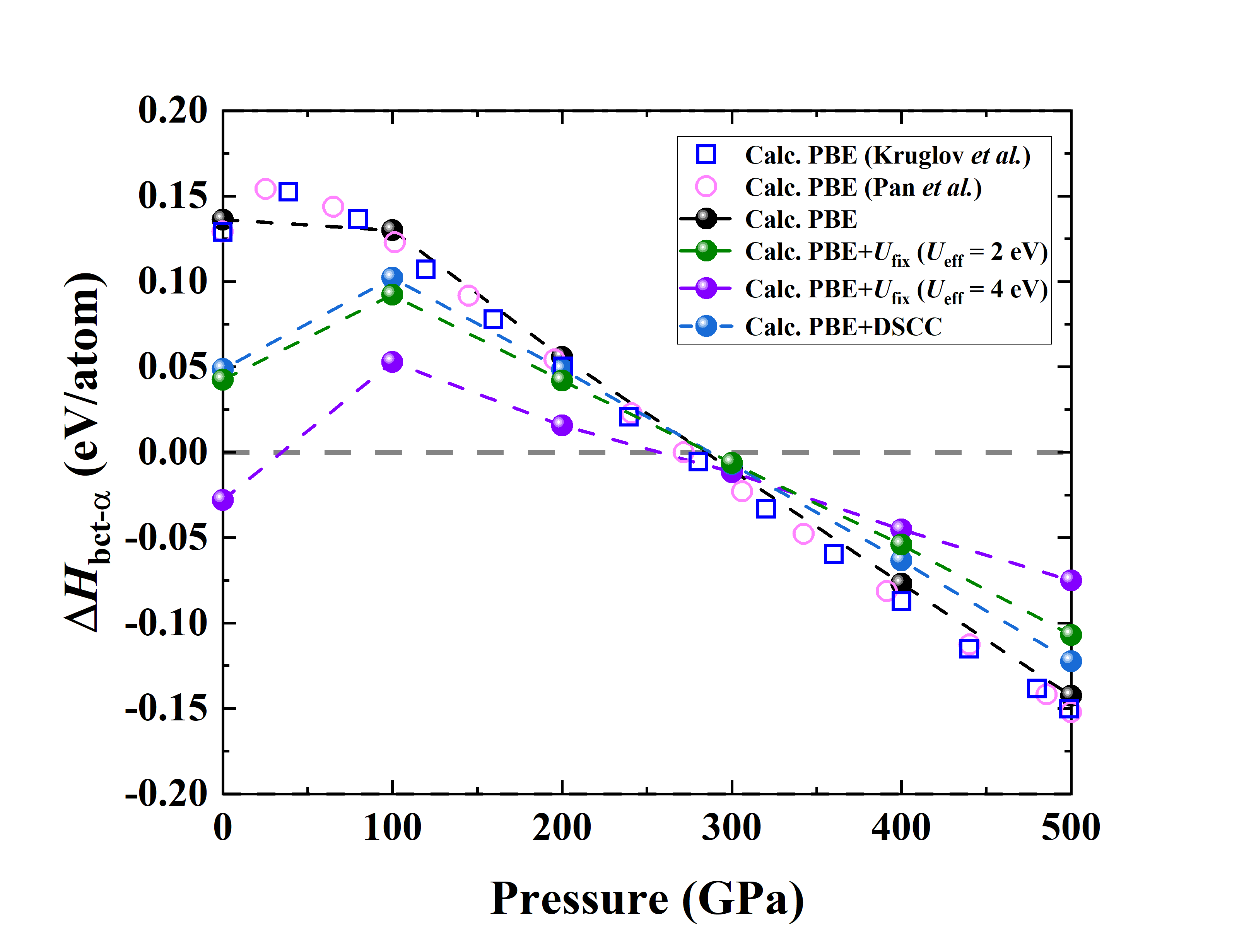}
\caption{The enthalpy difference of bct structure related to $\alpha$ phase of U as a function of pressure, calculated by PBE, PBE+$U_{\rm fix}$, and PBE+DSCC.}\label{fig:U-dh}
\end{figure}

The results of theoretical simulations of the enthalpy difference of bct-U relative to $\alpha$-U at different pressures is presented in Fig.\ref{fig:U-dh}. For all calculations, there is a transition from the $\alpha$-U to the bct-U at pressure about 270 GPa, which is confirmed by the previous theoretical works\cite{Adak2011,Kruglov2019,Pan2024}. The inclusion of $\Ueff$ almost does not change the pressure of the transition around 270 GPa. However, it reduces the enthalpy difference at 0 GPa. In particular, the PBE+$U_{\rm fix}$ with a large $\Ueff$ = 4 eV predicts that bct-U becomes the most stable structure at 0 GPa, which contradicts the experiments.

It should be noted that there is still some debate about the on-site Coulomb correction of uranium. The proposed scheme offers a more accurate compression behaviour, although it also overestimates the equilibrium volume at 0 GPa by $\sim$5\%. Furthermore, our calculations demonstrate that effective on-site Coulomb interaction strength $\Ueff$ = 4 eV results in a qualitatively incorrect ground-state phase, emphasising the importance of an appropriate on-site Coulomb correction. 

\subsection{Thorium}

\begin{figure}[tbp]
\centering
\includegraphics[width=0.49\textwidth]{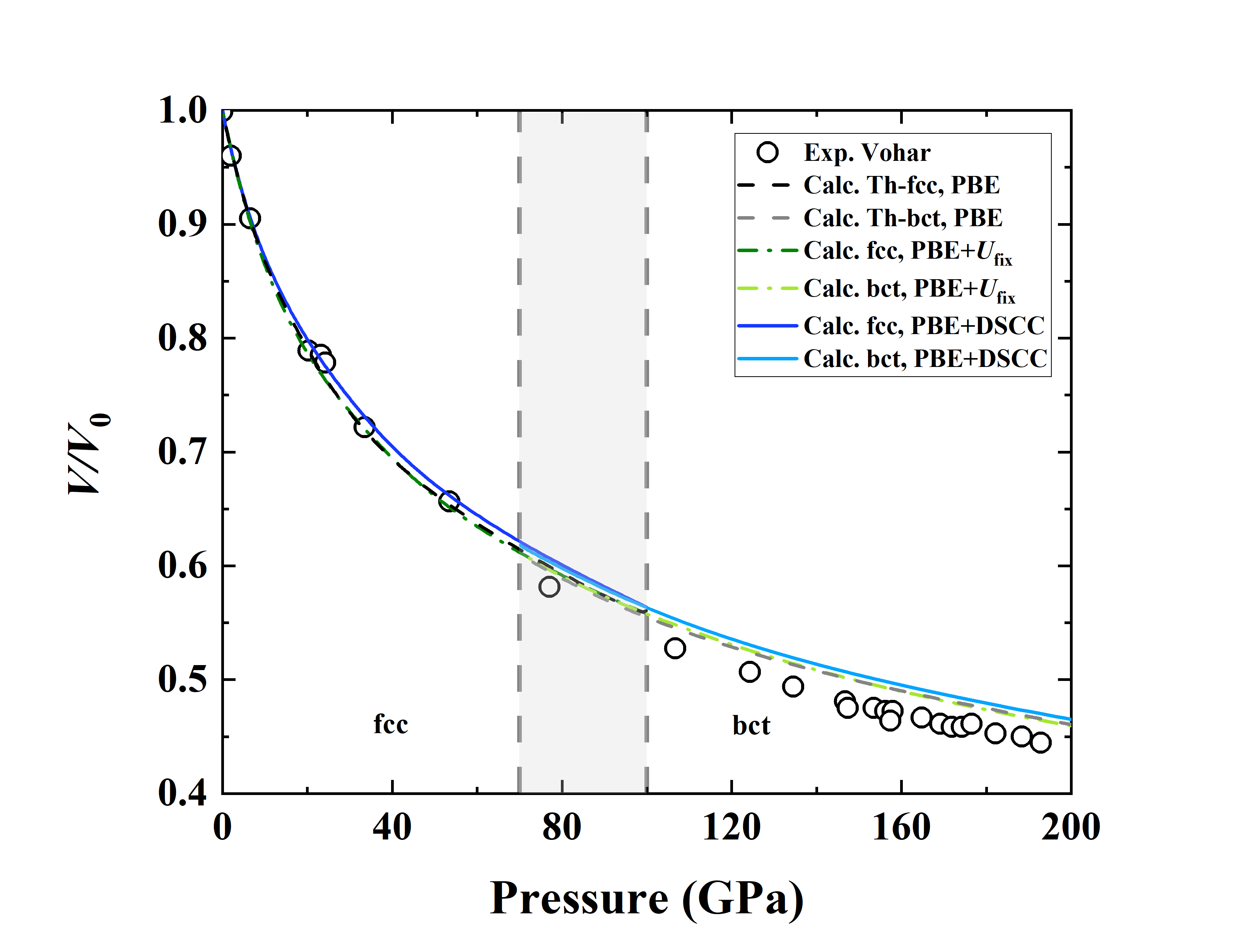}
\caption{The static EoS $P$-$V/V_0$ of Th calculated by PBE, PBE+$U_{\rm fix}$, and PBE+DSCC. Experimental data also present for comparison.}\label{fig:Th-eos}
\end{figure}

\begin{table}[tbp]
\caption{Structural parameters $V_0$, $B_0$ and $B'_0$ of Th by experimental and theoretical approaches.{\label{tab:struc:Th}}}
\begin{tabular*}{0.45\textwidth}{@{\extracolsep{\fill}} l|cccc}
\hline\hline
     &$V_0$ (\r{A}$^3$)          &$B_0$ (GPa)      &$B'_0$ \\
 \hline
Exp. Armstrong\cite{Armstrong1959}        &32.9     &58.0    &2.9        \\
Exp. Vohra\cite{Vohra1991}       &32.9     &55.0     &3.8 \\
PBE    &32.463     &45.100    &3.489          \\
PBE+$U_{\rm fix}$($\Ueff$=2 eV)   &34.035     &47.733     &4.223          \\
PBE+DSCC  &32.920     &49.132     &4.238          \\
\hline\hline
\end{tabular*}
\end{table}

Finally, the performance of the scheme on Th is examined, given that its correlation effect is relatively weak compared to other actinide metals. Indeed,  
PBE has performed relatively good in the description of Th \cite{Soderlind1995,Bouchet2006,Hu2010,Kyvala2020}. Whereas PBE+$U_{\rm fix}$ overestimates the equilibrium volume at 0 GPa (see Table \ref{tab:struc:Th}). Fig.\ref{fig:Th-eos} shows the static EoS for Th, PBE, PBE+$U_{\rm fix}$ and PBE+DSCC generally perform similarly, PBE being slightly closer to the experimental data. However, we will see in the following that PBE+$U_{\rm fix}$ may lead to an unreasonable prediction of the structure.

As Fig.\ref{fig:U-ueff} shows, At 0 GPa, the $\Ueff$ of Th is 1.5 eV. The on-site Coulomb interaction strength is less than those of uranium, which is consistent with the degree of $f$-electron localization in the actinide series \cite{Moore2009}. The $\Ueff$ of Th continues to decrease, with a trend similar to that of uranium. It decreases by about 45\% at 200 GPa.

\begin{figure}[tbp]
\centering
\includegraphics[width=0.49\textwidth]{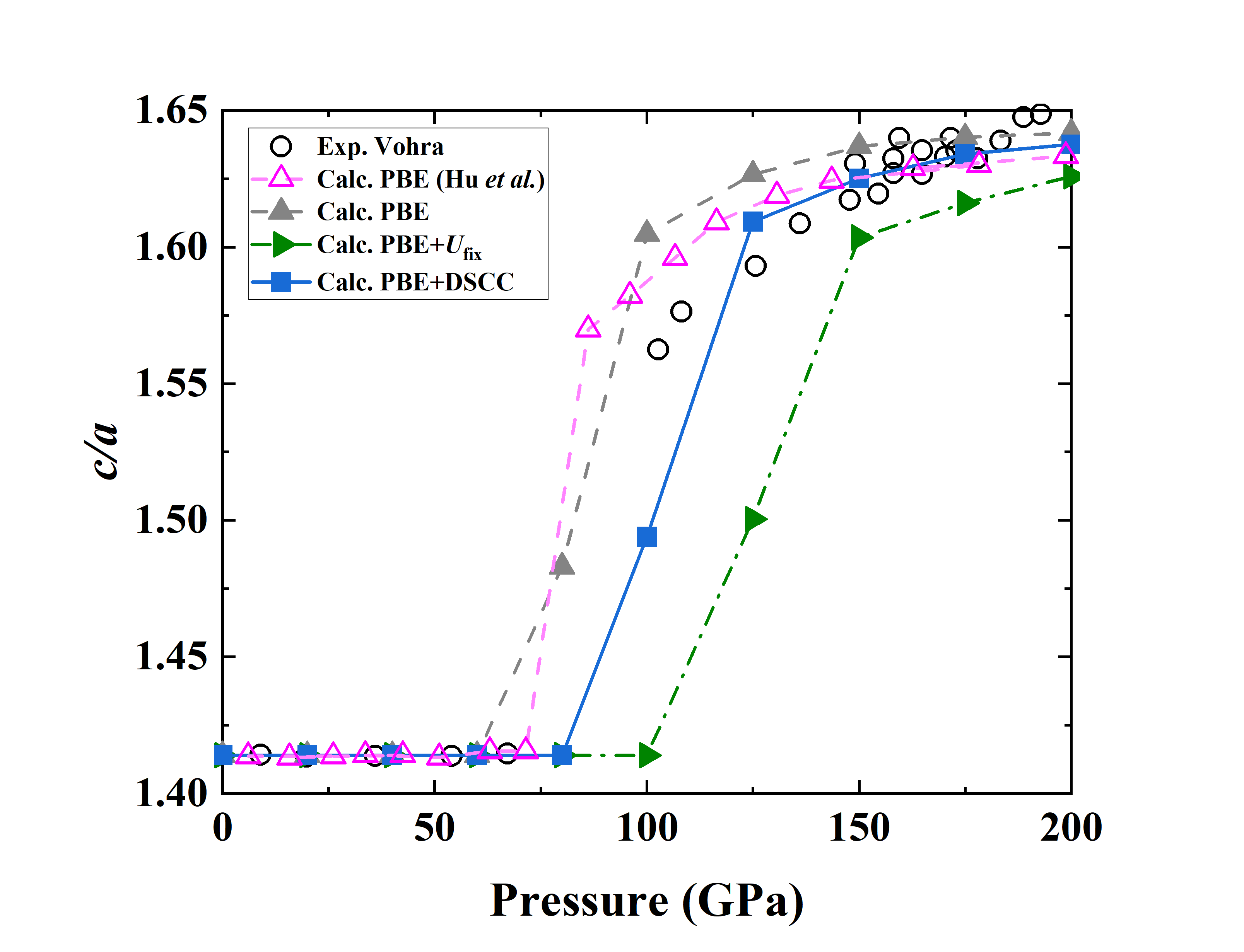}
\caption{The axial ratio ($c/a$) as a function of pressure of Th, calculated by PBE, PBE+$U_{\rm fix}$, and PBE+DSCC. The $c/a$ value of fcc structure is $\sqrt{2}$.}\label{fig:Th-dh}
\end{figure}

Within the pressure range of 70-100 GPa, Th undergoes a fcc-bct phase transition. Since the enthalpy difference between the two phases is too small \cite{Bouchet2006,Hu2010}, we investigate this phase transition in Fig.\ref{fig:Th-dh} by the axial ratio $c/a$ of the ground-state structure versus pressure. The axial ratio of the fcc structure is exactly $\sqrt{2}$. Below 60 GPa, PBE predicts a fcc structure. At 80 GPa, PBE yields the axial ratio $c/a$ of 1.48, which then increases rapidly to 1.60 at 100 GPa. Above 100 GPa, $c/a$ increases and reaches a plateau value of 1.64. Both PBE+$U_{\rm fix}$ and PBE+DSCC show a similar trend with PBE, but with a difference in the pressure that $c/a$ deviates from $\sqrt{2}$. PBE+DSCC scheme is capable of describing the fcc-bct phase transition up to 100 GPa. However, PBE+$U_{\rm fix}$ approach does not reproduce this transition until 100 GPa.

In the actinide series, an increase in correlation effect is observed with rising atomic number \cite{Moore2009}. This phenomenon can be reflected by the $\Ueff$ values of Th and U calculated by DSCC approach. In terms of structural properties, for Th, PBE has proven capable of providing a relatively good description. Our scheme can deal with systems with weak correlation effect, giving a satisfactory description of the EoS and phase stability. Conversely, PBE+$U_{\rm fix}$ with a larger fixed value of $\Ueff$ = 2 eV is unable to describe the fcc-bct phase transition within the 70-100 GPa.

%%%%%%%%%%%%%%%%%%%%%%%%%%%%%%%%%%%%%%%%%%%%%%%%%

\section{Conclusion and Perspectives}
\label{sec:Conclu}

In conclusion, we preformed calculations for typical lanthanides and actinides by means of an on-site Coulomb correction scheme, where the on-site Coulomb interaction strength is adaptively determined from the structure and component. 
The scheme reproduce good compressive properties over a wide range of pressures. It also provides a reasonable theoretical description of the phase stabilities. 
The calculated results emphasise the essential role of the pressure-dependent on-site Coulomb interaction in describing of the structural behaviour of $f$-electron systems. Our recently developed DSCC method can be used as a suitable strategy to determine the pressure-dependent on-site Coulomb interaction strength self-consistently. 
In addition, the simulated on-site Coulomb interaction decreases under compression, which is consistent with the picture of $f$-electron delocalization.
The discrepancies between the results of our scheme and experiments may be mainly caused by the absence of subtle many-body effects such as the hybridization of $f$-$spd$ electrons and the associated Kondo physics, which are hardly captured by the current static mean-field based approach. These deficiencies will be hopefully removed by more sophisticated DMFT calculations using on-site Coulomb interaction parameters obtained by DSCC.

\section{Acknowledgement}

We thank Yu Liu,  Yuan-Ji Xu,  Hua-Jie Chen, Ming-Feng Tian for helpful discussions. This work was supported by the National Key Research and Development Program of China (Grant No. 2021YFB3501503), the National Natural Science Foundation of China (Grant Nos. U2230401, U1930401, U23A20537) and Funding of National Key Laboratory of Computational Physics. We thank the Tianhe platforms at the National Supercomputer Center in Tianjin.

\small

\bibliographystyle{unsrt}

\bibliography{bib}

\end{document}